\documentclass[conference]{IEEEtran}
\IEEEoverridecommandlockouts
\usepackage{cite}
\usepackage{amsmath,amssymb,amsfonts}
\usepackage{algorithmic}
\usepackage{graphicx}
\usepackage{textcomp}
\usepackage{xcolor}
\usepackage{braket}
\usepackage{subcaption}
\usepackage{tikz}
\usetikzlibrary{quantikz}

\def\BibTeX{{\rm B\kern-.05em{\sc i\kern-.025em b}\kern-.08em
    T\kern-.1667em\lower.7ex\hbox{E}\kern-.125emX}}

\begin{document}

\title{Advanced Quantum Poisson Solver in the NISQ era}

\author{\IEEEauthorblockN{1\textsuperscript{st} Walter Robson}
\IEEEauthorblockA{\textit{Department of Computer Science and Engineering} \\
\textit{University Of Notre Dame}\\
Notre Dame, IN 46556 USA \\
wrobson@nd.edu}
\and
\IEEEauthorblockN{2\textsuperscript{nd} Kamal K. Saha}
\IEEEauthorblockA{\textit{Center for Research Computing} \\
\textit{University Of Notre Dame}\\
Notre Dame, IN 46556 USA \\
ksaha@nd.edu}
\and
\IEEEauthorblockN{3\textsuperscript{rd} Connor Howington}
\IEEEauthorblockA{\textit{Center for Research Computing} \\
\textit{University Of Notre Dame}\\
Notre Dame, IN 46556 USA \\
chowingt@nd.edu}
\and
\IEEEauthorblockN{4\textsuperscript{th} In-Saeng Suh}
\IEEEauthorblockA{\textit{~~~ National Center for Computational Sciences} \\
\textit{Oak Ridge National Laboratory}\\
Oak Ridge, TN 37380 USA \\
suhi@ornl.gov}
\and
\IEEEauthorblockN{5\textsuperscript{th} Jaroslaw Nabrzyski}
\IEEEauthorblockA{\textit{Center for Research Computing} \\
\textit{University Of Notre Dame}\\
Notre Dame, IN 46556 USA \\
naber@nd.edu}


}

\maketitle

\begin{abstract}
The Poisson equation has many applications across the broad areas of science and engineering. 
Most quantum algorithms for the Poisson solver presented so far, either suffer from lack of accuracy and/or 
are limited to very small sizes of the problem, and thus have no practical usage.
Here we present an advanced quantum algorithm for solving the Poisson equation with high accuracy 
and dynamically tunable problem size. After converting the Poisson equation to the linear systems 
through the finite difference method, we adopt the Harrow-Hassidim-Lloyd (HHL) algorithm as the basic framework. 
Particularly, in this work we present an advanced circuit that ensures the accuracy of the solution 
by implementing non-truncated eigenvalues through eigenvalue amplification as well as by increasing the accuracy 
of the controlled rotation angular coefficients, which are the critical factors in the HHL algorithm. 
We show that our algorithm not only increases the accuracy of the solutions, but also composes more practical and 
scalable circuits by dynamically controlling problem size in the NISQ devices. 
We present both simulated and experimental solutions, and conclude that overall results on the quantum hardware are dominated by the error in the CNOT gates.
\end{abstract}


\section{Introduction}

The Poisson equation is a second order partial differential equations widely used in various fields of science and engineering.
In general, in order to solve the Poisson equation numerically, projection methods including such as collocation, spectral, 
and boundary element methods as well as finite-difference methods\cite{b1} are used.
The core of these methods is to approximate the solution of the Poisson equation as the solution of linear systems. 
However, since the dimension of the linear system obtained from the discrete Poisson equation is generally very large, 
this belongs to the class of problems demanding much computational time.

Therefore, a fascinating technology that would significantly reduce the computational cost of solving the Poisson equation
is the application of quantum computing which is faster and more powerful computation than classical computing.
Cao et al.\cite{cao} firstly used the HHL algorithm\cite{hhl} to solve the Poisson equation with the quantum circuit model. 
Later, Wang et al.\cite{wang} pointed out a bottleneck of Cao's algorithm in implementing the controlled rotation of HHL.
In their work, they developed new way of implementing the controlled rotation of HHL using named quantum function-value binary expansion (qFBE)\cite{borwein} and quantum algorithms for solving the reciprocal and square root operations based on the classical non-restoring method.
Thus, they not only reduced the algorithm's complexity, but also made the circuit complete and implementable.
However, even though their works apparently improved the quantum algorithm and circuits for the Poisson solver,
these still either suffer from lack of accuracy and/or are limited to very small sizes of the problem, and thus have no practical usage.

Here we present an advanced quantum algorithm for solving the Poisson equation with high accuracy and dynamically tunable problem size. 
Particularly, we develop a circuit based on the Wang's algorithm that ensures the accuracy of the solution
by implementing non-truncated eigenvalues through eigenvalue amplification as well as by increasing the accuracy
of the controlled rotation angular coefficients, which are the critical factors in the HHL algorithm.

As a result, we achieve higher success probability which typically decreases with a rate of $1/\kappa^2$, 
where $\kappa$ is proportional to the number of finite difference discretization, as the problem size is increased.
We show that our algorithm not only increases the accuracy of the solutions, but also composes more practical and
scalable circuits by dynamically controlling problem size in the NISQ devices.
We present both simulated and experimental results, and discuss the sources of errors.
Finally, we discuss how the overall results on the currently available quantum hardware are dominated by the error in the CNOT gates.

\section{Overview of the problem}

The goal of this work is to implement a quantum algorithm solving the multi-dimensional Poisson equation with boundary conditions.  
Let us consider the Poisson equation defined in an open bounded domain $\Omega \subset \Re^d$, where $d$ is the number of spatial dimensions.
\begin{align}
	-\nabla^2 v(x) = b(x), \ x \ \text{in} \ \Omega \ \
    \\
	v(x) = 0,  \ x \ \text{on} \ \delta\Omega \ \ \Omega = (0,1)^d
\end{align}
where $\delta \Omega$ is the boundary of $\Omega$. 
One way to solve this problem is to discretize $\Omega$ in $N^\prime = N + 1$ grid points in each dimension, where $N$ is an exponent of base $2$ in this work. 
The solution $v(x)$ is a vector of $(N-1)^d$ entries. 

In this work, we focus on the one-dimensional Poisson equation with Dirichlet boundary conditions. Using the central-difference approximation 
to discretize the second-order derivative, Eq.\,(1) could be converted to finite difference form as

\begin{eqnarray}
	A \cdot
    \begin{pmatrix}
    v_1 \\ v_2 \\ \vdots \\ v_{N-1}
    \end{pmatrix} 
	 & = & 
	\frac{1}{h^{2}}
    \begin{pmatrix}
     2      & -1      &        &  0     \\
    -1      & \ddots  & \ddots &        \\
            & \ddots  & \ddots & -1     \\
     0      &         & -1     &  2 
    \end{pmatrix}
    \cdot
    \begin{pmatrix}
    v_1 \\ v_2 \\ \vdots \\ v_{N-1}
    \end{pmatrix}   \nonumber \\ 
	 &=&  
    \begin{pmatrix}
    b_1 \\ b_2  \\ \vdots \\ b_{N-1}
    \end{pmatrix}
\end{eqnarray}

We now have the $N-1$ linear equation system, $i.e.$, $A \ket{v} = \ket{b}$ to be solved.
Matrix $A$ is a Hermitian matrix and has the dimension of $(N-1) \times (N-1)$ and the mesh size $h$ equals to $1/N$. 
The best numerical algorithms for solving this problem run polynomially with matrix size\cite{shewchuk} in classical computing, 
so the run-time increases exponentially with the dimension of the problem. 
In this paper a quantum algorithm is used to produce a quantum state representing the normalized solution of the problem. 
Since this technique runs in polylog time the curse of dimensionality can be broken.
Thus, we will now solve the linear system of equations based on the HHL algorithm\cite{hhl}.

\section{Quantum Algorithm and Circuit Design}

\begin{figure}
    \centering
    \includegraphics[scale=.49]{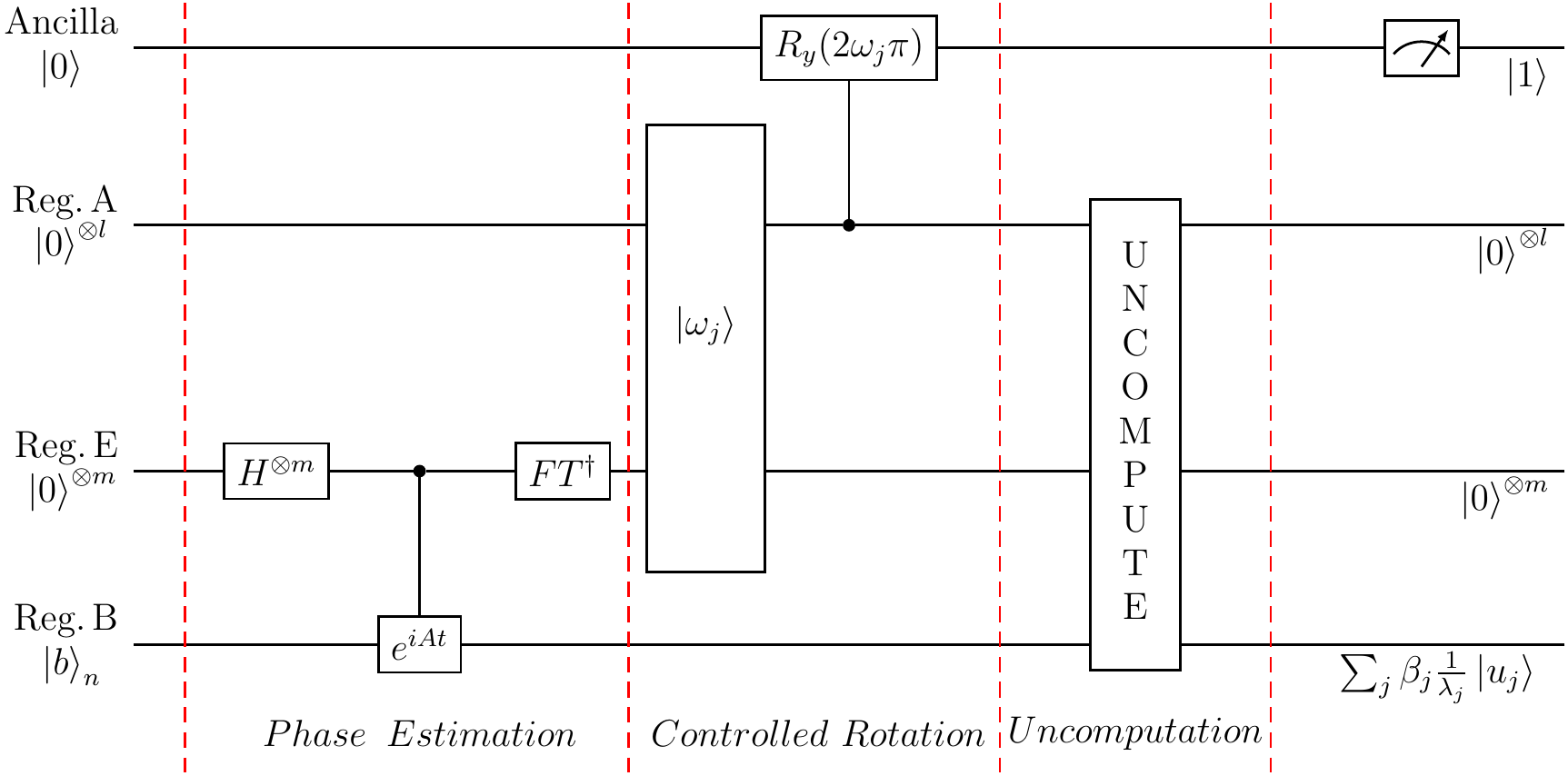}
	\caption{The overall circuit representation of the algorithm for solving one-dimensional Poisson equation. The number of qubits of registers A, E and B is $l$, $m$ and $n$, respectively. $|\omega_j\rangle$ is the angular coefficient evolved from the approximated eigenvalue $|\lambda_j\rangle$, output of the QPE. The input  
	$|b\rangle_n = \sum_{i=1}^{2^n-1} b_i|i\rangle$ is prepared and stored in register B.}
\label{fig1}
\end{figure}

The overall circuit diagram of the present algorithm for solving the one-dimensional Poisson equation is presented in Fig.\,1. As the figure shows, the algorithm consists of the phase estimation, the controlled rotation, 
and the uncomputation stages. Its circuit diagram has three main registers -- reg.\,B, reg.\,E, and reg.\,A. Reg.\,B is used to encode the coefficients of the right-hand side of Eq.\,(1) and its number of qubits is $n=\lceil \log(N^\prime) \rceil$.
Reg.\,E is used to store the approximated eigenvalues and its number of qubits is $m=2n + 2 + f$, where the most significant $2n + 2$ bits hold the integer part and the remaining $f$ bits the fractional part. Reg.\,A is used to store calculated angular coefficients for the controlled rotation operation, and its number of qubits is chosen to be $l \ge m$. 

In this work, we assume that the input state $\ket{b}$ of reg.\,B is prepared as \(\sum_i b_i|i\rangle \), 
where \(b_i\) is the value on the right-hand side of Eq.\,(3), and \(|i\rangle\)  is the computational basis\cite{aaronson}.  
That is, the input $\ket{b}$ assigns the pre-requisite state-vector. In other words, reg.\,B contains the problem that we are trying to solve, which we then entangle with the approximated eigenvalues 
\({\lambda}_j\) on reg.\,E. 
The output of the algorithm thus is a quantum state representing the solutions of the Poisson equation as probability amplitudes on reg.\,B.  
Therefore, this circuit is a process of quantum state preparation, with the output written as: 
\(|v\rangle = A^{-1}|b\rangle = \sum_i\alpha_i|i\rangle \), where \(\alpha_i\) is the value of the solutions of the Poisson equation
after normalization.

The algorithm used in this work follows several steps:
\begin{itemize}
	\item Prepare the quantum state $\sum_{i}b_i \ket{i}$ in reg.\,B.

	\item Use Quantum Phase Estimation (QPE) algorithm on regs.\,B and E. This algorithm applies several Hamiltonian simulations of $U = e^{i A t}$ with $t = 2\pi \frac{1}{2^n} 2^k$, \, $k=0,..., n-1$, to reg.\,B and entangles the eigenvalues $\lambda_j$ of matrix $A$ in reg.\,E with the eigenstates $\ket{u_j}$ in reg.\,B. The system has now the state: $\sum_{j} \beta_j  \ket{{\lambda}_j} \ket{u_j}$.
    
	\item Apply the controlled rotation which consists of two parts: preparing the rotation angular coefficients $\ket{\omega_j}$ in reg.\,A and performing the controlled $R_y$ operation on the ancillary quibit.
	
	\item Uncompute QPE and $\ket{\omega_j}$ operations on regs.\,A, E and B.
    
	\item Measure the ancillary qubit. If the measurement of the qubit results in state $\ket{1}$, the algorithm successfully transforms reg.\,B into the solution $\sum_{j} \beta_j \frac{1}{{\lambda}_j} \ket{u_j}$. Otherwise the algorithm has to be restarted.
\end{itemize}

\begin{figure}
\centering
\includegraphics[scale=.28]{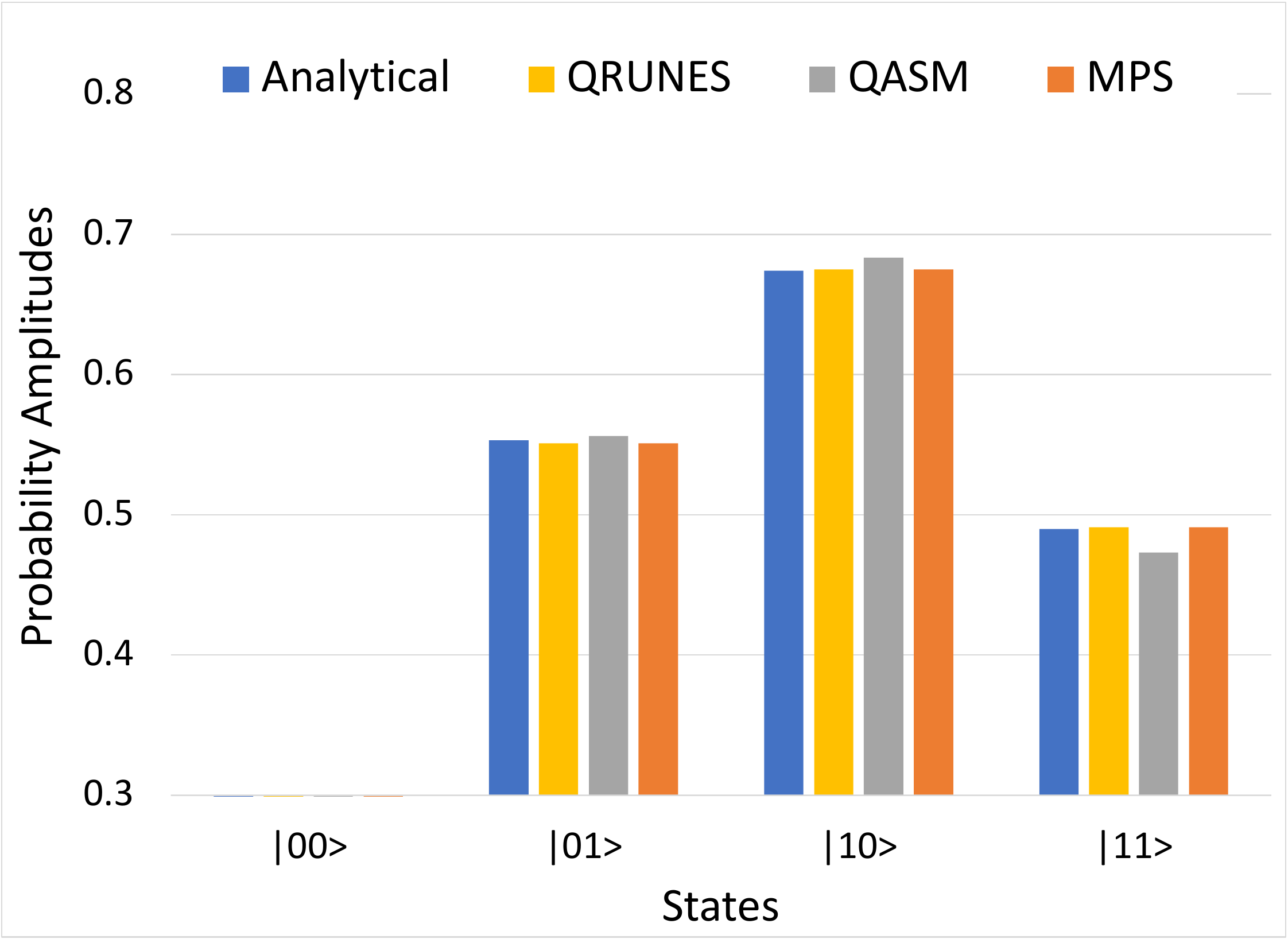}
\includegraphics[scale=.28]{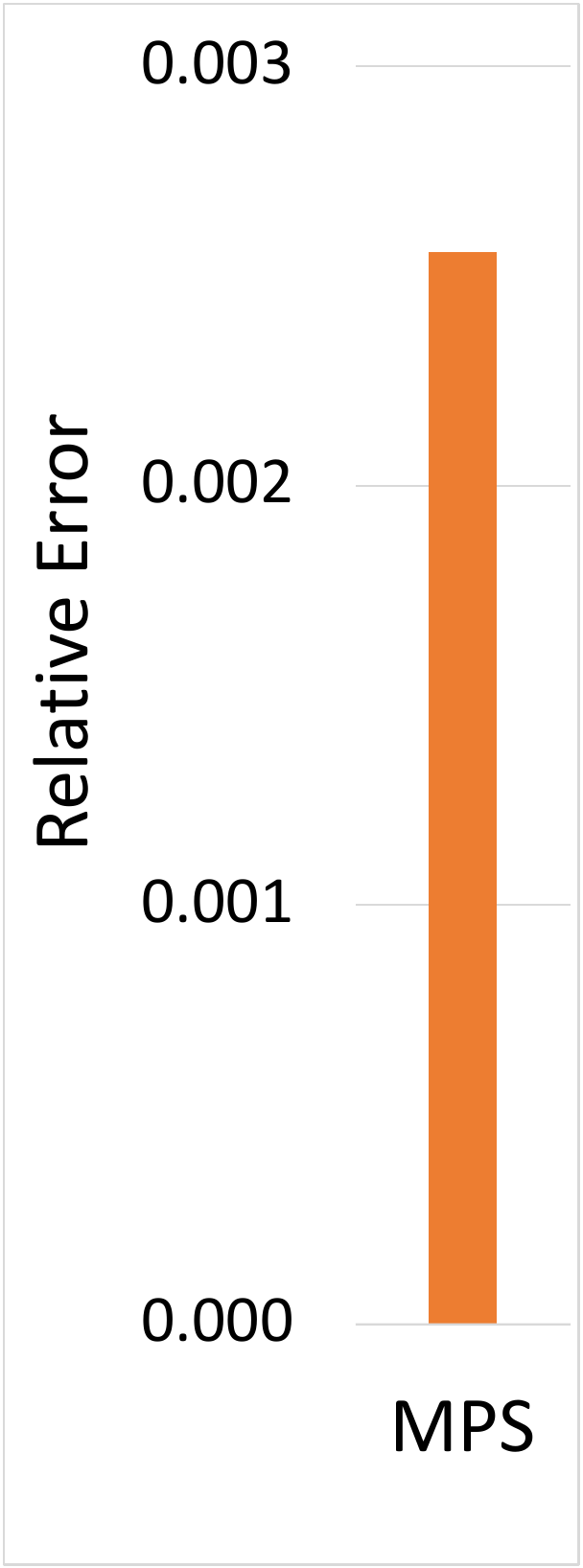}
        \caption{(Left) IBM's QASM and MPS-based simulated Poisson equation solution comparing with analytical and existing QRUNES\cite{wang} results for $3\times 3$ problem size. (Right) Shows relative error in MPS-based solution with respect to the analytical result.}
\label{fig2}
\end{figure}

We now discuss a few key steps of the algorithm as follows:
\subsection{Quantum Phase Estimation}

Phase estimation is used to approximate the eigenvalues of the discretized matrix $A$
and entangle the states encoding the eigenvalues with the corresponding eigenstates\cite{luis}. 
Hamiltonian simulation of $e^{i A t}$ is the crucial part in phase estimation. 
Therefore, we first start with exploiting properties of matrix $A$ to efficiently solve the HHL algorithm by simulating 
the unitary operator $e^{i A t}$. 
The eigenvalues of matrix $A$ are $\lambda_j = 4 N^2 {\rm sin}^2(j \pi/2 N)$ and its corresponding eigenvectors are
$u_{j}(k) = \sqrt{2/N} {\rm sin}(j \pi k/N)$\cite{demmel}.
Utilizing the properties of matrix $A$, the unitary operator can be decomposed with a Hermitian matrix $S$ ($S$ being an
orthogonal matrix composed of the eigenvectors of $A$), and finally can be diagonalized via the sine transform. After sine transform and phase kickback, we adopt a quantum algorithm for the square root operation based on the classical non-storing method\cite{sutikno}. The detailed circuit composition for phase estimation is discussed in Ref.\cite{wang,saha}.

\subsection{Controlled Rotation}

After phase estimation, we perform the linear map taking the state of $\ket{{\lambda}_j}$ to $(1/\lambda_j) \ket{\lambda_j}$.
This process consists of two parts: calculating the rotation angular coefficients and 
performing the controlled $R_y$ operation. The probability amplitude of $1 / {\lambda}_j$ would be produced by implementing the
controlled $R_y$ rotation, that is, $R_y (2 \theta_j) \ket{0} = \cos{\theta_j} \ket{0} + \sin{\theta_j} \ket{1}$, 
where the rotation angle $\theta_j$ can be expressed in terms of of  ${\lambda}_j$ as 
\begin{equation} \label{}
		\sin{\theta_j} = 1/{\lambda}_j
\end{equation}
Which can be rewritten as,
\begin{equation} \label{}
	\cot{\theta_j} = \sqrt{{\lambda}^{2}_j - 1}, ~~~ \theta_j \in (0, \pi/2)
\end{equation}
Taking $\theta_j = \omega_j \pi$, then Eq.\,(5) becomes
\begin{equation} \label{}
	\omega_j = \frac{1}{\pi} {\rm arccot} \big(\sqrt{{\lambda}^{2}_j - 1} \big), ~~~ \omega_j \in (0, 1/2)
\end{equation}

where $\omega_j$ is the rotation angular coefficient. Preparing them through the circuit design has an exponential cost with the problem size\cite{wang}, so in our implementation we pre-prepare $\omega_j$ and encode them into the circuit.

\begin{figure}
    \centering
	\includegraphics[scale=.28]{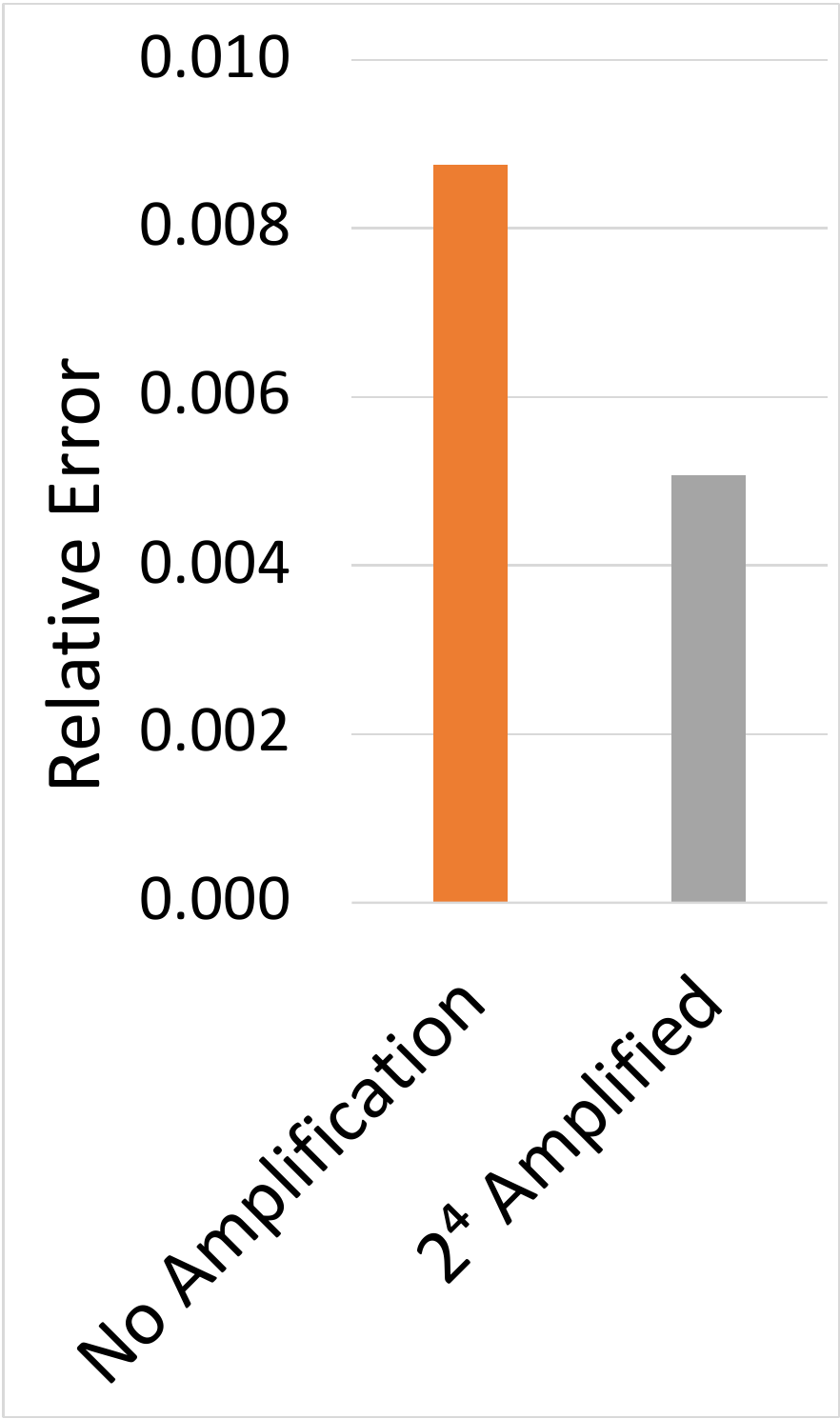}
    \includegraphics[scale=.28]{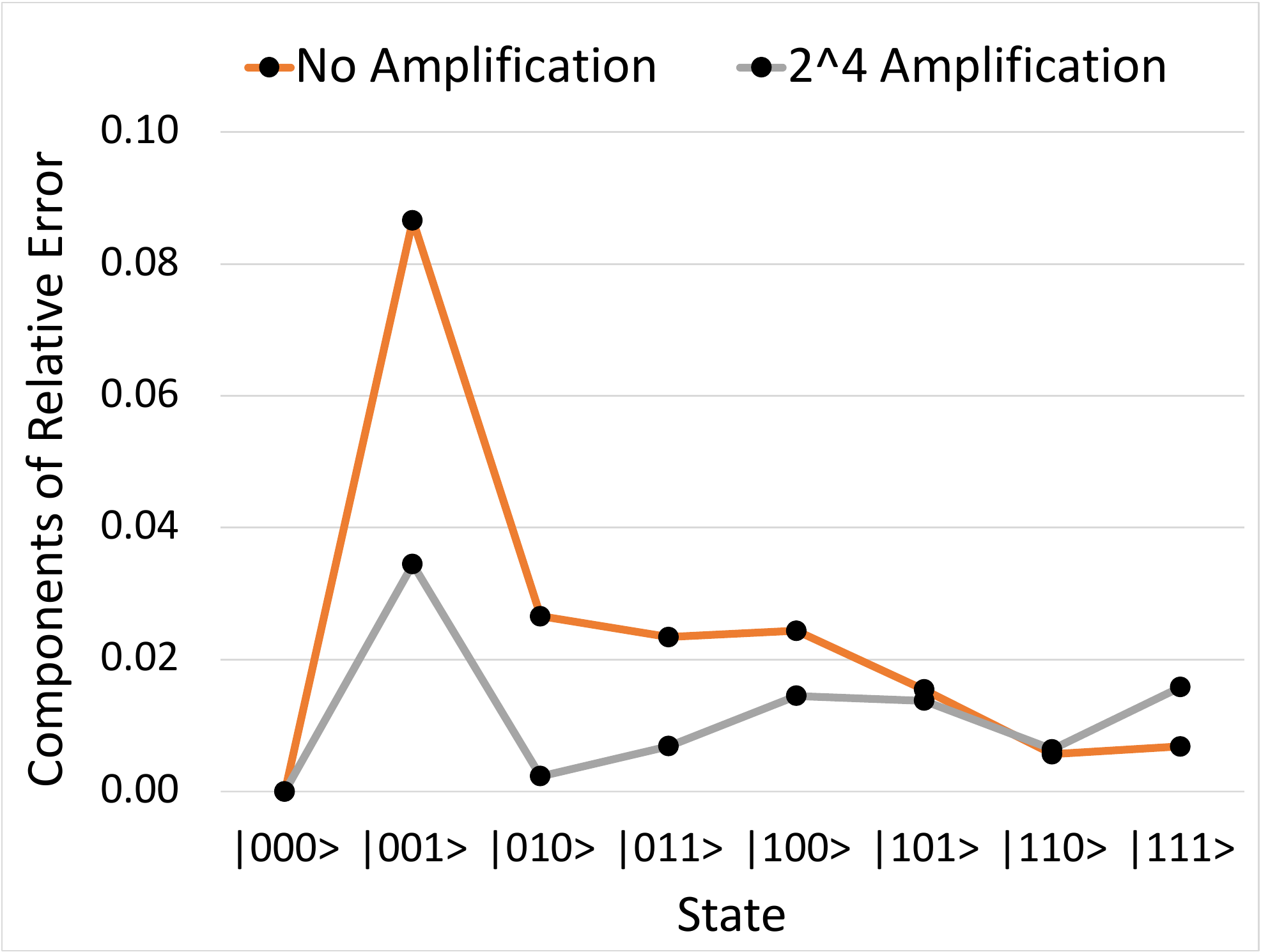}
        \caption{(Left) Shows the improvement in the relative error to the solution simulated with MPS for a $7\times 7$ problem. (Right) Explicitly shows those relative errors at the level of quantum states.}
\label{fig4}
\end{figure}

\subsection{Eigenvalue Amplification}

After controlled rotation, the uncomputation is implemented to evolve the state of register B, E, and A back to the initial state. Finally, we perform the measurement operation. If the measurement result of the ancillary qubit is $\ket{1}$, then we get the solution state for the Poisson equation.

However, as the problem size is increased, the success probability of obtaining the desired state in one computation is decreased with a rate of $1/\kappa^2$, where $\kappa$ is proportional to the number of finite difference discretization. While expanding the problem size, we find that errors in the eigenvalues start to accumulate in the calculation of the rotation angular coefficient and thus identify a critical source of error in the truncation of the eigenvalues of the matrix $A$.

Therefore, in order to improve the accuracy of the algorithm, we amplify the eigenvalue by a factor of $2^i$ ($i$ is an integer) in QPE and making sure that it is being compensated by a normalization factor in the later section of the circuit\cite{saha}. 
This amplification is critical to getting accurate results from the larger matrices,
as the success rate was so low that the number of times required to run the circuit quickly becomes a limiting factor. 
However, the eigenvalue amplification significantly improves the accuracy of the of the computation resulting in a improving the success probability.

\section{Results and Discussions}

We first demonstrate our algorithm and circuit with IBM's QASM and MPS simulators by producing the result for relatively a smaller problem defined by the size of the matrix $A$ in Eq.\,(3). The solution of a $3 \times 3$ problem with $\ket{b} = 0.0 |00\rangle + \frac{1}{\sqrt{2}}|01\rangle + \frac{1}{2}(|10\rangle + |11\rangle)$ being the right-hand side of the Poisson equation, is shown in Fig.\,2 and that also includes the analytical solution for comparison. Though there is some discrepancies of QASM-based result, the MPS-based solution shows reasonable agreement with both analytical and existing QRUNES\cite{wang} results. The accuracy of the MPS-based result is depicted through the relative error in MPS solution with respect to the analytical result, shown in the right side of Fig.\,2. 

The implementation of the circuit for a $3 \times 3$  problem requires 32 qubits which is exactly the maximum limit of QASM simulator. In contrast, MPS simulator allows up to 100 qubits. Therefore, as a viable option for presenting the circuit of larger problem sizes, we decided to proceed with the MPS simulator only. The result for a $7 \times 7$ problem with $\ket{b} = 0.0 |000\rangle + \frac{1}{4}(|001\rangle + |010\rangle + |011\rangle + |100\rangle) + \frac{1}{2}(|101\rangle + |110\rangle + |111\rangle)$ is shown in Fig.\,3. With the implementation of the eigenvalue amplification by a factor of $2^4$ reduces the relative error by about half, and we are confident that a higher amplification factor (for example, $2^8$) would further reduce this error. The right part of this figure explicitly shows the relative error at the level of the quantum states. 

Finally, for a $3 \times 3$ problem, we experiment our circuit on the IBM Brooklyn 65 qubit system (Hummingbird r2) and present the results along with MPS-based solution in Fig.\,4. The circuit compiled with different level of transpiler optimization parameter and sampling\cite{younis} is used, yet the experimental results come out as erroneous with an artifact of nonzero contribution for $|00\rangle$ state. The average CNOT error on Brooklyn system is $8.094\text{e-}2$, in other words they have an accuracy of about $0.92$. Thus, we can estimate the total accuracy of the experiment based on the final number of CNOT gates transpiled from the more abstract circuit, approximated by ${0.92}^c$ where $c$ is the final number of CNOT gates after transpilation. Our optimized circuit has about 5.5k CNOT gate and their accumulated error results in washing out the accuracy of the overall experimental solution. 

\begin{figure}
    \centering
    \includegraphics[scale=.31]{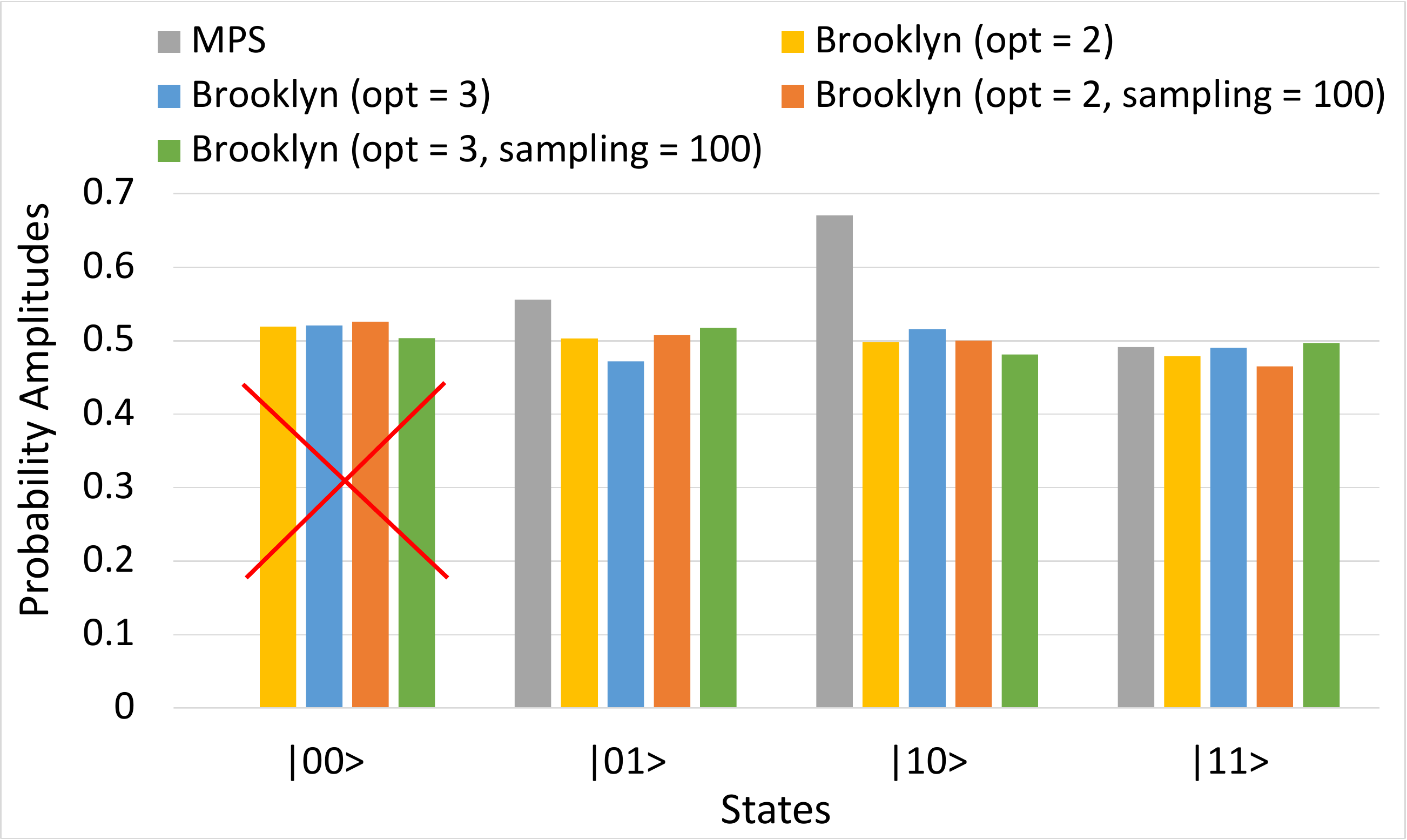}
        \caption{For a $3 \times 3$ problem, comparison of the experimental results, pursued on the $ibm_{-}brooklyn$ quantum device, with MPS-based solution. Experiment was done on optimized circuits compiled with different levels of transpiler optimization parameters and sampling. The nonzero contribution to the $\ket{00}$ state marked with cross is an experimental artifact due to the hardware error involving CNOT gates.} 
\label{fig5}
\end{figure}

\section*{Acknowledgment}
This research was supported in part by the University of Notre Dame's Center for Research Computing. This manuscript has been authored by UT-Battelle, LLC under Contract No. DE-AC05-00OR22725 with the U.S. Department of Energy. The United States Government retains and the publisher, by accepting the article for publication, 
acknowledges that the United States Government retains a non-exclusive, paid up, irrevocable, world-wide license to publish or reproduce the published form of the manuscript, or allow others to do so, for United States Government purposes. 
The Department of Energy will provide public access to these results of federally sponsored research in accordance 
with the DOE Public Access Plan (http://energy.gov/downloads/doe-public-access-plan). 
This research used resources of the Oak Ridge Leadership Computing Facility at the Oak Ridge National Laboratory, 
which is supported by the Office of Science of the U.S. Department of Energy under Contract No. DE-AC05-00OR22725.







\begin{thebibliography}{00}
\bibitem{b1} Poisson equation, numerical methods. Encyclopedia of Mathematics. URL: 
	$http://encyclopediaofmath.org/index.php?title=Poisson_{-}equation,_{-}numerical_{-}methods\&oldid=48217$
\bibitem{cao} Y. Cao, A. Papageorgiou, I. Petras, J. Traub, and S. Kais, ``Quantum algorithm and circuit design solving the Poisson equation,'' New J. of Phys, vol. 15, p. 013021, 2013.
\bibitem{hhl} A. W. Harrow, A. Hassidim and S. Lloyd, ''Quantum algorithm for solving linear systems of equations,'' Phys. Rev. Lett., 103, 150502, 2009.
\bibitem{wang} S. Wang, Z. Wang, W. Li, L. Fan, Z. Wei, and Y. Gu, ``Quantum fast Poisson solver: the algorithm and complete and modular circuit design,'' Quantum Info. Processing, vol. 19, p. 170, 2020.
\bibitem{borwein} J. M. Borwein, R. Girgensohn, ``Addition theorems and binary expansions,'' Can. J. Math., 47, 262, 1995
\bibitem{shewchuk} J. Shewchuk, ''An introduction to the conjugate gradient method without the agonizing pain,'' 1994.
\bibitem{aaronson} S. Aaronson, ``Read the fine print,'' Nat. Phys. 11, 291, 2015.
\bibitem{luis} A. Luis and J. Pe\ifmmode \check{r}\else \v{r}\fi{}ina, ``Optimum phase-shift estimation and the quantum description of the phase difference,'' Phys. Rev. A 54, 4564, 1996.
\bibitem{demmel} J. W. Demmel, {\it Appliied numerical linear algebra} (SIAM, Philadelphia, 1997), Chap. 6.
\bibitem{sutikno} T. Sutikno, ``An efficient implementation of the non-restoring square root algorithm in gate level,'' Int. J. Comput. Theory Eng., 3, 46, 2011.
\bibitem{saha} K. K. Saha, W. Robson, C. Howington, I. Suh, and J. Nabrzyski, ``Advanced Quantum Poisson Solver in the NISQ era: Scaling Performance and Error Analysis,'' unpublished 2022.
\bibitem{younis} E. Younis, C. Iancu, ``Quantum Circuit Optimization and Transpilation via Parameterized Circuit Instantiation,'' arXiv:2206.07885, 2022.  
\end{thebibliography}
\end{document}